\documentclass[runningheads]{llncs}
\usepackage{graphicx}
\usepackage{float}
\usepackage{hyperref}

\usepackage{booktabs} 
\usepackage{amsmath}
\usepackage{amssymb}
\usepackage{xcolor}

\begin{document}
\title{SAMVAD: A Multi-Agent System for Simulating Judicial Deliberation Dynamics in India}
\titlerunning{SAMVAD: Simulating Judicial Deliberation}
\author{Prathamesh Devadiga\inst{1}\orcidID{0009-0000-2948-4799} \and
Omkaar Jayadev Shetty\inst{1}\orcidID{0009-0006-5874-8822} \and
Pooja Agarwal\inst{1}\orcidID{0009-0000-8553-4711}}
\authorrunning{P. Devadiga et al.}
\institute{PES University, Bengaluru, India \\
\email{\{pes2ug22cs410, pes2ug22cs377\}@pesu.pes.edu}, \email{poojaagarwal@pes.edu}}
\maketitle
\begin{abstract}
Understanding the complexities of judicial deliberation is crucial for assessing the efficacy and fairness of a justice system. However, empirical studies of judicial panels are constrained by significant ethical and practical barriers. This paper introduces SAMVAD, an innovative Multi-Agent System (MAS) designed to simulate the deliberation process within the framework of the Indian justice system. Our system comprises agents representing key judicial roles—a Judge, a Prosecution Counsel, a Defense Counsel, and multiple Adjudicators (simulating a judicial bench)—all powered by large language models (LLMs). A primary contribution of this work is the integration of Retrieval-Augmented Generation (RAG) grounded in a domain-specific knowledge base of landmark Indian legal documents, including the Indian Penal Code and the Constitution of India. This RAG functionality enables the Judge and Counsel agents to generate legally sound instructions and arguments, complete with source citations, thereby enhancing simulation fidelity and transparency. Adjudicator agents engage in iterative deliberation rounds, processing case facts, legal instructions, and arguments to reach a consensus-based verdict. We detail the system architecture, agent communication protocols, the RAG pipeline, the simulation workflow, and a comprehensive evaluation plan designed to assess performance, deliberation quality, and outcome consistency. This work provides a configurable and explainable MAS platform for exploring legal reasoning and group decision-making dynamics in judicial simulations, specifically tailored to the Indian legal context and augmented with verifiable legal grounding via RAG.

\keywords{Multi-Agent Systems \and Agent-Based Simulation \and Legal AI \and Judicial Deliberation \and Retrieval-Augmented Generation (RAG) \and LLM Agents \and Indian Law.}
\end{abstract}
\section{Introduction}
The rapid advancement of artificial intelligence, particularly in the era of big data, presents a profound opportunity to transform the judicial sector. Integrating AI into the administration of justice is no longer a futuristic concept but an emerging imperative for enhancing efficiency and consistency \cite{ref_gptjudge}. For over four decades, AI research has explored legal applications, evolving from early judicial expert systems to more sophisticated models of legal reasoning \cite{ref_judicialreasoning}. However, a significant challenge remains in simulating the nuanced cognitive processes of judges and judicial panels. This study aims to address this gap by developing an intelligent system that models judicial thinking, thereby promoting judicial progress, reducing costs, and improving case-handling efficiency.

Judicial deliberation is a complex, interactive process involving legal reasoning, argumentation, and the application of abstract legal principles to specific facts. Computationally modeling this process offers a powerful tool for studying group decision-making, analyzing the impact of evidence presentation, aiding in legal training, and testing the capabilities of AI in complex reasoning tasks. Traditional approaches in legal AI have often focused on static outcome prediction or rule-based expert systems, which fail to capture the dynamic, dialogic nature of deliberation.

Multi-Agent Systems (MAS) offer a natural paradigm for modeling such distributed, interactive systems. By simulating autonomous agents with distinct roles and behaviors, MAS can reveal emergent properties of group dynamics. The recent advent of Large Language Models (LLMs) further empowers these agents with sophisticated reasoning and natural language capabilities. However, a critical limitation of general-purpose LLMs is their potential for factual inaccuracies ("hallucinations") and their lack of deep, domain-specific knowledge unless specifically augmented.

This paper introduces SAMVAD (Simulated Agent-based Multi-agent Verdict Adjudication), an MAS framework that overcomes these limitations by integrating Retrieval-Augmented Generation (RAG). We construct a domain-specific legal knowledge base from foundational Indian legal texts, such as the Indian Penal Code (IPC) and the Constitution of India. This knowledge base grounds the reasoning of our Judge and Counsel agents, enabling them to generate legally precise instructions and arguments. Crucially, these agents are designed to cite their sources, providing a mechanism for explainability and ensuring their outputs are anchored in authoritative legal texts.

Our key contributions are:
\begin{itemize}
    \item An innovative MAS architecture designed to model the Indian judicial deliberation process, featuring distinct Judge, Counsel, and Adjudicator agents.
    \item The integration of a RAG pipeline with a vector knowledge base derived from Indian legal documents to ground the reasoning of key agents.
    \item A citation mechanism that allows RAG-enhanced agents to reference the source documents used in their generated outputs, promoting transparency and verifiability.
    \item A robust orchestration framework that manages the simulation workflow, including multi-round deliberation and consensus checking.
\end{itemize}

\section{Related Work}
The intersection of AI and law has evolved from early expert systems to modern machine learning models for tasks like document analysis and outcome prediction. The emergence of LLMs has opened new frontiers, enabling more dynamic and interactive simulations of legal processes.

Recent work such as SimuCourt has demonstrated the potential of agent-based systems to simulate various court proceedings, testing an "Agent-as-Judge" across different case types \cite{ref_agentcourt}. These systems aim to automate tasks like precedent search, case analysis, and judgment generation. Similarly, other frameworks have utilized LLM-based agents to represent judges and jurors, exploring the dynamics of a collaborative judicial process \cite{ref_agentsbench}. In these models, judge agents typically moderate discussions and ensure adherence to legal standards, while juror agents contribute perspectives based on community values. The prompts for these agents are often engineered to reflect their distinct roles, with judges focusing on legal principles and jurors on societal considerations.

Our work builds on these foundations but is distinct in several critical aspects. While general social simulations have widely adopted agent-based modeling to study emergent phenomena like opinion dynamics, few have specifically targeted the structured and role-based nature of formal legal deliberation. Furthermore, traditional computational law models often rely on symbolic logic or argumentation frameworks, which lack the fluidity of natural language interaction that LLMs provide.

The core innovation of SAMVAD is its deep integration of a domain-specific RAG pipeline grounded in Indian law. LLM-powered "agentic workflows" are becoming increasingly common, where agents perceive a state, reason using an LLM, and generate an action. However, without grounding, these agents are prone to error in knowledge-intensive domains like law. By augmenting our Judge and Counsel agents with a verifiable knowledge base, we address this critical limitation. Unlike general legal benchmarks such as LegalBench-RAG, which evaluate RAG on static tasks \cite{ref_legalbenchrag}, our system employs RAG dynamically within an interactive simulation. This ensures that the generated legal instructions and arguments are not only contextually relevant but also legally sound and citable, enhancing both the fidelity and explainability of the simulation. To our knowledge, SAMVAD is one of the first LLM-based MAS frameworks to simulate judicial deliberation with an integrated, domain-specific RAG for legal grounding, particularly within the context of the Indian legal system.

\section{Methodology}
Our simulation framework is a modular Python system composed of an Orchestrator, specialized Agent types, a Legal Knowledge Base with RAG capabilities, and an LLM Interface. The Orchestrator manages the simulation lifecycle, coordinating interactions between the Judge, Counsel, and Adjudicator agents. The system's legal reasoning is anchored in a knowledge base derived from authoritative Indian legal documents, which are made accessible via a semantic search-enabled vector database.

\subsection{Architecture Design}
\begin{figure}[H]
    \centering
    \includegraphics[width=0.9\textwidth]{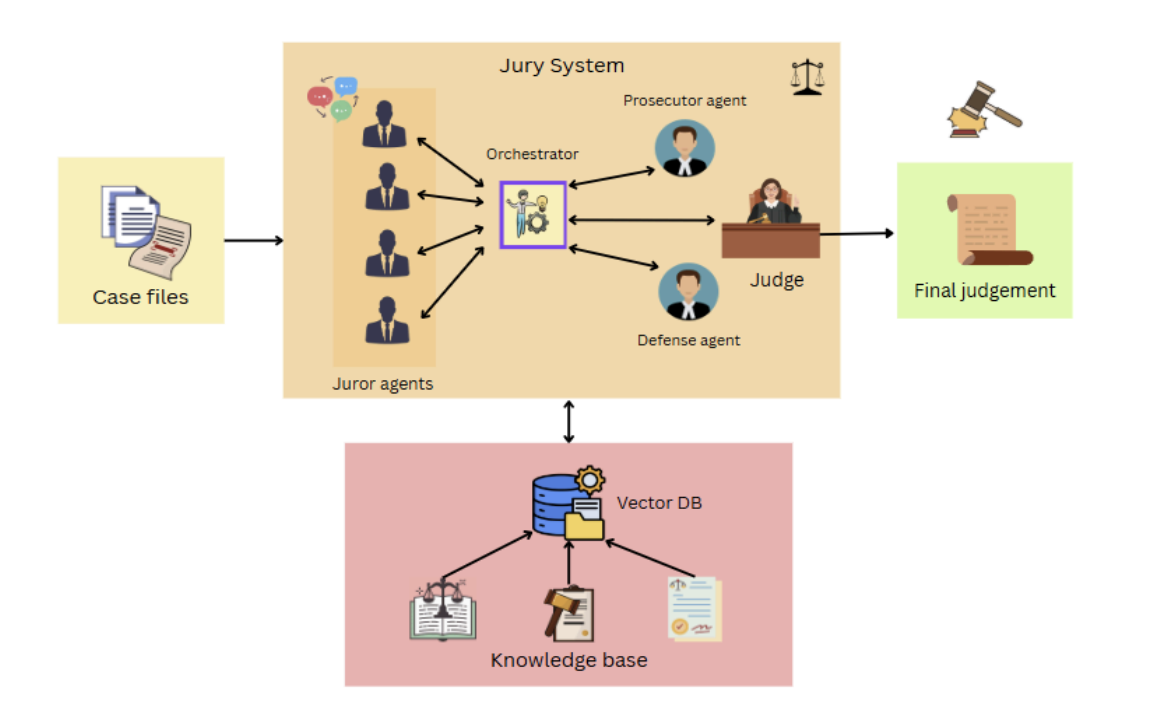}
    \caption{High-level architecture of the SAMVAD system. The Orchestrator manages the interaction between agents (Judge, Counsel, Adjudicators), who leverage a RAG pipeline connected to a legal knowledge base to process case files and produce a final judgment.}
    \label{fig:architecture}
\end{figure}

The system architecture, depicted in Figure \ref{fig:architecture}, is designed for modularity and clarity. The Orchestrator acts as the central hub, managing the flow of information and the sequence of actions. Case files, structured in JSON format, serve as the primary input. Each case file includes a unique \texttt{case\_id}
, a factual summary, the specific legal charges, a simplified explanation of the relevant laws, itemized lists of prosecution and defense evidence, and a set of keywords for grounding the analysis. This structured input ensures that all agents begin with a consistent and comprehensive understanding of the case.

\subsection{Agents}
\subsubsection{Adjudicator Agents.}
These agents form the core of the deliberative body, simulating a judicial bench. Each Adjudicator agent operates autonomously, analyzing the case summary, legal instructions from the Judge, and arguments from both Counsels. In each deliberation round, they formulate and articulate their current leaning (Guilty, Not Guilty, or Undecided) along with a detailed justification. Their reasoning evolves as they process the arguments of their peers, contributing to the collective decision-making process and making the simulation a realistic reflection of group deliberation.

\subsubsection{Counsel Agents (Prosecution \& Defense).}
The Counsel agents are responsible for constructing legally grounded arguments for their respective sides. These agents are enhanced with RAG capabilities. Upon receiving the case file, they query the legal knowledge base to retrieve relevant statutes, definitions, and precedents that support their position. This allows them to generate arguments that are not only persuasive but also legally precise, complete with citations to the source material, thereby increasing the realism and depth of the simulation.

\subsubsection{Judge Agent.}
The Judge agent is a specialized agent tasked with generating impartial instructions for the Adjudicator agents. Using RAG, it analyzes the case summary and charges to retrieve the relevant legal principles and standards of proof from the vectorized knowledge base. Its prompt is carefully engineered to synthesize this retrieved context with the case facts, resulting in clear, accurate, and citable instructions that frame the deliberation for the Adjudicators.

\subsection{Orchestrator}
The Orchestrator is the central controller that coordinates the multi-agent simulation, mimicking the procedural flow of a judicial proceeding. It initializes the simulation, distributes the case file, and sequences the agent interactions: first the Judge's instructions, followed by the arguments from the Prosecution and Defense Counsels. It then manages the iterative deliberation rounds among the Adjudicators, collecting their statements and checking for consensus after each round. The Orchestrator also logs key performance metrics, such as agent response times and reliance on external knowledge. If consensus is not reached within a predefined number of rounds, the Orchestrator can declare a hung panel and conclude the simulation.

\subsection{Knowledge Base and RAG}
The foundation of our system's legal fidelity is its knowledge base, constructed from foundational Indian legal texts.
\begin{itemize}
    \item \textbf{The Constitution of India:} Provides the supreme legal framework, defining fundamental rights, state obligations, and the powers of the judiciary.
    \item \textbf{The Indian Penal Code (IPC):} Serves as the official criminal code of India, defining substantive criminal offenses and their corresponding punishments.
    \item \textbf{The Code of Criminal Procedure (CrPC):} Outlines the procedural law for the administration of criminal justice, from investigation to trial and sentencing.
\end{itemize}

To create the RAG pipeline, these documents are processed into a vector store. The text from each PDF is extracted and segmented into manageable chunks, retaining metadata like the source document. A sentence-embedding model converts these chunks into dense vector representations, which capture their semantic meaning. These embeddings are stored in a persistent ChromaDB vector database. When a RAG-enabled agent (Judge or Counsel) needs to perform a task, its query is converted into an embedding and used to perform a similarity search against the vector database, retrieving the most relevant legal text chunks to serve as context for the LLM's generation.

\section{Simulation Workflow}
\begin{figure}[H]
    \centering
    \includegraphics[width=\textwidth]{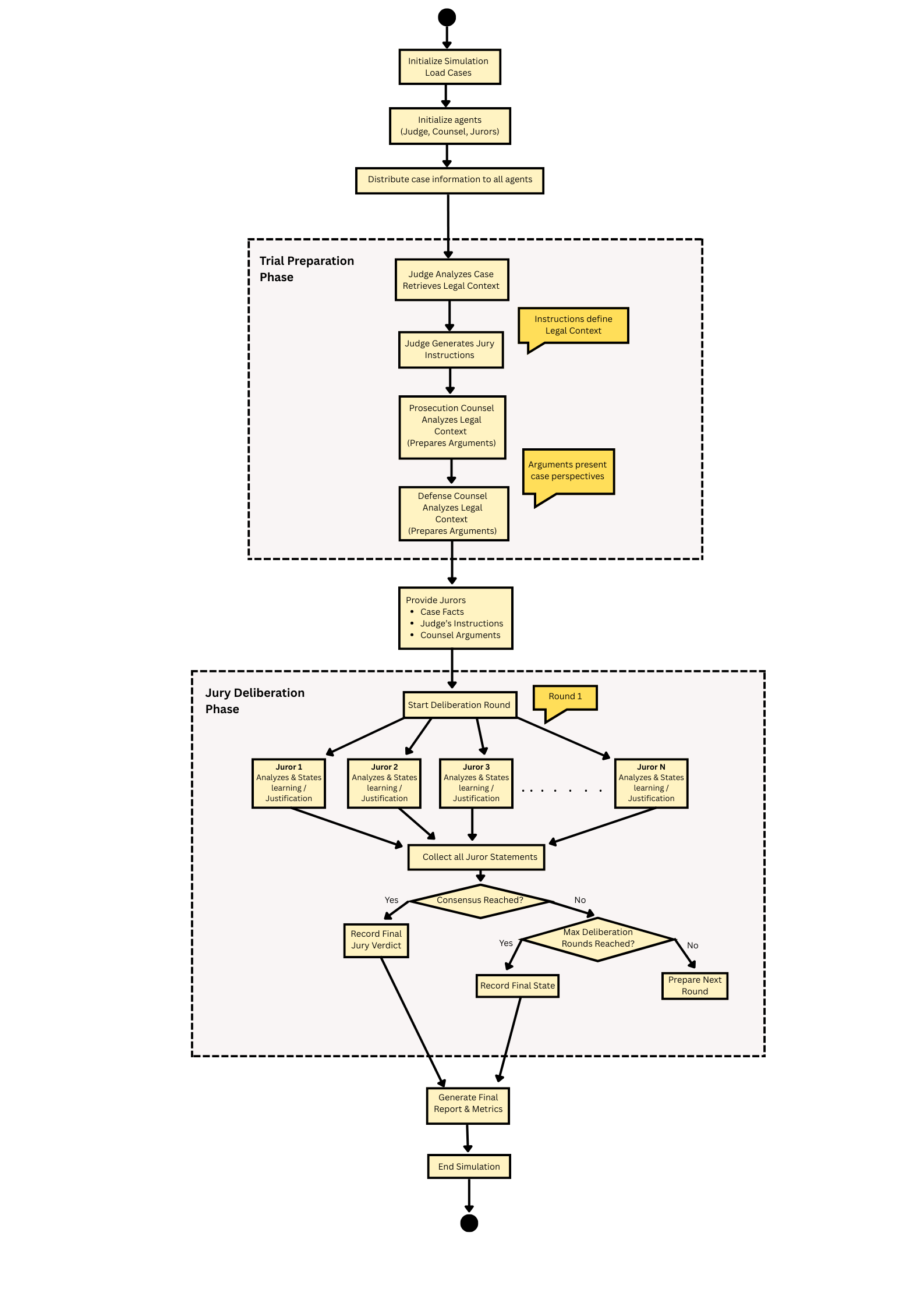}
    \caption{Detailed Simulation Workflow, illustrating the trial preparation and iterative jury deliberation phases.}
    \label{fig:workflow}
\end{figure}

The simulation proceeds through a structured sequence of phases, as illustrated in Figure \ref{fig:workflow}.
\begin{enumerate}
    \item \textbf{Initialization:} The system environment is set up. The legal knowledge base is loaded from the persistent vector store or created anew if not present. The specific case file is loaded, and the agents (Judge, Counsels, Adjudicators) are instantiated with their designated LLMs and roles.
    \item \textbf{Trial Preparation Phase:} Case information is distributed to all agents. The Judge agent uses RAG to query the knowledge base and generate impartial instructions for the adjudicators. Concurrently, the Prosecution and Defense Counsel agents use RAG to retrieve legal context and prepare their arguments, grounding their claims in the knowledge base.
    \item \textbf{Jury Deliberation Phase:} The deliberation begins. In each round, the Adjudicator agents analyze the case facts, instructions, arguments, and the statements from their peers. They then produce their individual leaning and a justification.
    \item \textbf{Consensus Check:} After each round, the Orchestrator collects all statements and checks if a verdict consensus (e.g., >80\% agreement) has been reached.
    \item \textbf{Conclusion:} If consensus is achieved, the final verdict is recorded. If not, the deliberation proceeds to the next round, up to a maximum limit. The simulation concludes by generating a final report detailing the outcome, deliberation history, and performance metrics.
\end{enumerate}

\section{Evaluation and Results}
We evaluate the system using a suite of metrics focused on performance, deliberation quality, and outcome consistency.

\subsection{Evaluation Metrics}
\begin{itemize}
    \item \textbf{Performance Metrics:} Measure the computational efficiency of the system, primarily through the response times of LLM calls (mean, median, min, max). This helps assess system latency.
    \item \textbf{Participation and Communication Metrics:} Quantify the engagement of Adjudicator agents. This includes the total number of statements and the proportion of "meaningful" statements (those with well-supported justifications). The participation rate tracks the percentage of agents contributing in each round.
    \item \textbf{Argument Grounding Score:} Measures the quality of an agent's reasoning by evaluating how well its justifications align with key case facts and legal principles. This is calculated by checking the frequency of predefined case keywords in agent statements. A higher score indicates a more fact-based deliberation.
    \item \textbf{Consistency Measures:} Assess the reproducibility of verdicts across multiple runs of the same case. We measure the verdict consistency rate (the percentage of runs yielding the most common verdict) and the distribution of different verdicts.
\end{itemize}

\subsection{Simulation Results}
Table \ref{tab:results} presents a summary of metrics from five simulated case runs, each with five adjudicators and with RAG enabled for the Judge and Counsel agents. The results show a high degree of participation (100\% across all cases) and strong consensus, with final agreement ratios of 0.90 or higher. Cases 1, 2, and 5 reached a verdict in a single round, indicating clear-cut scenarios, while Cases 3 and 4 required two rounds, suggesting more complex deliberations. The Argument Grounding Score varied, with Case 3 showing the highest score (0.45), indicating a deliberation that was more closely tied to the provided case facts.

\begin{table}[H]
\centering
\caption{Simulation Results for Five Sample Cases.}
\label{tab:results}
\begin{tabular}{@{}lccccc@{}}
\toprule
\textbf{Metric} & \textbf{Case 1} & \textbf{Case 2} & \textbf{Case 3} & \textbf{Case 4} & \textbf{Case 5} \\ \midrule
No. of Adjudicators & 5 & 5 & 5 & 5 & 5 \\
RAG enabled (Judge) & True & True & True & True & True \\
RAG enabled (Counsel) & True & True & True & True & True \\
Final Verdict & Not Guilty & Not Guilty & Guilty & Not Guilty & Guilty \\
Deliberation Rounds & 1 & 1 & 2 & 2 & 1 \\
Final Agreement Ratio & 1.00 & 1.00 & 0.90 & 1.00 & 0.95 \\
Adjudicator Participation Rate & 1.00 & 1.00 & 1.00 & 1.00 & 1.00 \\
Avg. Meaningful Statements & 1.00 & 1.00 & 1.20 & 1.15 & 1.10 \\
per Adjudicator & & & & & \\
Avg. Argument Grounding Score & 0.32 & 0.30 & 0.45 & 0.42 & 0.38 \\
\bottomrule
\end{tabular}
\end{table}

\subsection{Ablation Studies}
To isolate the impact of the RAG component and LLM choice, we conducted ablation studies. We ran simulations for a fixed case with and without RAG enabled for the Judge and Counsel agents, across four different open-source LLMs. As shown in Table \ref{tab:ablation_llm_rag}, the results demonstrate a clear and consistent positive impact from RAG. Across all models, enabling RAG led to higher agreement ratios, significantly improved argument grounding scores (roughly doubling them), more meaningful statements, and higher verdict consistency. For instance, with the Qwen-2.5-7B model, enabling RAG increased the grounding score from 0.21 to 0.42 and improved verdict consistency from "Medium" to "Very High." This strongly suggests that grounding agent reasoning in an external, verifiable legal knowledge base is critical for achieving high-quality, consistent, and explainable simulation outcomes.

\begin{table}[H]
\centering
\caption{Ablation Study: Impact of RAG and Model Choice on Simulation Metrics.}
\label{tab:ablation_llm_rag}
\setlength{\tabcolsep}{4pt}
\begin{tabular}{@{}lcccccc@{}}
\toprule
\textbf{Model} & \textbf{RAG (Judge)} & \textbf{RAG (Counsel)} & \textbf{Agreement} & \textbf{Ground Score} & \textbf{Avg. Stmts.} & \textbf{Consistency} \\
\midrule
Qwen-2.5 7B & Yes & Yes & 0.98 & 0.42 & 1.15 & Very High \\
Qwen-2.5 7B & No  & No  & 0.90 & 0.21 & 0.95 & Medium \\
LLaMA-3 8B  & Yes & Yes & 0.95 & 0.37 & 1.05 & High \\
LLaMA-3 8B  & No  & No  & 0.87 & 0.18 & 0.85 & Low \\
Mistral 7B  & Yes & Yes & 0.96 & 0.39 & 1.10 & High \\
Mistral 7B  & No  & No  & 0.89 & 0.20 & 0.90 & Medium \\
Gemma 7B    & Yes & Yes & 0.94 & 0.35 & 1.00 & High \\
Gemma 7B    & No  & No  & 0.86 & 0.17 & 0.80 & Low \\
\bottomrule
\end{tabular}
\end{table}

\end{document}